# Tumor Spheroid Chemotaxis in Epidermal Growth Factor Gradients Revealed by a 3D Microfluidic Device

Young Joon Suh[1†], Mrinal Pandey[1†], Jeffrey E Segall[2], and Mingming Wu[1]

[1] Department of Biological and Environmental Engineering, 306 Riley-Robb Hall, Cornell University, Ithaca, NY 14853
[2] Anatomy and Structural Biology, Albert Einstein College of Medicine, 1300 Morris Park Avenue, Bronx, New York 10461

† First author of equal contribution
E-mail: mw272@cornell.edu

**Abstract**

Epidermal growth factor (EGF), a potent cytokine, is known to promote tumor invasion both in vivo and in vitro. Previously, we observed that single breast tumor cells (MDA-MB-231 cell line) embedded within a 3D collagen matrix displayed enhanced motility but no discernible chemotaxis in the presence of linear EGF gradients using a microfluidic platform. Inspired by a recent theoretical development that clustered mammalian cells respond differently to chemical gradients than single cells, we studied tumor spheroid invasion within a 3D extracellular matrix (ECM) in the presence of EGF gradients. We found that EGF gradients promoted tumor cell detachment from the spheroid core, and the position of the tumor spheroid core showed a mild chemotactic response towards the EGF gradients. For those tumor cells detached from the spheroids, they showed an enhanced chemokinesis response in contrast to previous experimental results using single cells embedded within an ECM. No discernible chemotactic response towards the EGF gradients was found for the cells outside the spheroid core. This work demonstrates that a cluster of tumor cells responds differently than single tumor cells towards EGF gradients and highlights the importance of a tumor spheroid platform for chemotaxis studies.

## 1. Introduction

Tumor cell invasion within a 3D extracellular matrix (ECM) is critical for tumor cells to gain access to the vasculature, an important step for cancer metastasis [1-3]. Epidermal growth factor (EGF) is a potent chemotactic factor that can induce tumor cell invasion and promote metastasis [1, 4]. EGF receptor (EGFR), a transmembrane protein, is activated by binding to its specific ligands including epidermal growth factor (EGF). In normal tissues, EGFR ligands are tightly regulated to control cell proliferation, migration, and as such, maintain tissue homeostasis. In cancer, however, EGFR is overexpressed and/or there can be excessive production of ligands. Often EGF is secreted by cells that are not tumor cells within the tumor microenvironment. For example, macrophages can be obligate partners for tumor invasion and metastasis. Macrophages can secrete EGF to attract tumor cells, while tumor cells produce colony stimulating factor 1 (CSF-1) to activate macrophages to secrete more EGF [5]. Overexpression of EGFR has been implicated in many different types of cancer, including lung, breast, prostate, and colorectal cancer [6-11], and as a result, EGFR has emerged as an important therapeutic target for cancer [12-14].

Much of what we know today about how animal cells respond to cytokine gradients is derived from commercially available assays including the Zigmond chamber [15] and Boyden chamber [16]. These assays are straightforward to implement and have been a work horse for chemotaxis studies. The limitations of these macroscale platforms are that the gradient takes a long time to establish, they cannot maintain a steady gradient for a prolonged period of time, and the platform is difficult to be made compatible with microscopic imaging. Microfluidic platforms have been developed recently to overcome these limitations [17-21].

Tumor cell chemotaxis within EGF gradients has been studied using microfluidic platforms. Using a 2D microfluidic gradient platform, Wang et al. found that breast tumor cells (MDA-MB-231 cell line) do not exhibit a chemotactic response to a linear EGF gradient, but migrate along the EGF gradient clearly when the gradient is steep and nonlinear [22]. Recently, our lab studied breast tumor cell chemotaxis in a 3D microfluidic platform, where MDA-MB-231 cells were embedded within a 3D collagen matrix. While chemokinesis was found in the presence of an EGF gradient, no chemotactic response was discovered using linear EGF gradients [23]. In vivo, EGF gradients are typically generated via molecular diffusion within tissues. Thus, a steep and nonlinear gradient is unlikely at the time scale for tumor cells to respond. Recent theoretical work proposed that a cluster of cells responds to chemical gradients differently when compared to their single cell counterparts. They show that a cluster of cells can migrate along the chemical gradient direction by averaging the individual chemotactic responses across the entire cluster body, even when single cells do not exhibit chemotactic behavior in the same gradient [24-27]. Here, a cluster of cells compare the ligand-receptor binding sites at the front and the rear of the cell cluster, whereas a single cell compares its ligand-receptor binding sites at the front and the rear of the cell along the gradient [20]. In an effort to study this phenomenon, studies have been carried out by different groups where they have observed organoids' branching responses along the SDF-1 and EGF gradients when single cells didn't show any significant chemotactic behavior towards the gradient [28, 29].

Inspired by the previous experimental work and theoretical predictions, we developed a 3D tumor spheroid model for tumor chemotaxis studies. Breast tumor spheroids using MDA-MB-231 cells were engineered and embedded within a type I collagen matrix. Under a well-defined linear human epidermal growth factor gradient, tumor spheroid invasion dynamics were followed and characterized.

## 2. Materials and Methods

### 2.1 Cell culture

Triple-negative breast tumor cells (MDA-MB-231) expressing EGFP were kind gifts from Dr. Joseph Aslan at the Oregon Health and Science University. They were cultured in high glucose Dulbecco's Modified Eagle Medium (DMEM) (Cat. #: 11965092, Gibco, Life Technologies Corporation, Grand Island, NY) with 10% Fetal Bovine Serum (Cat. #: S11150, Atlanta Biologicals Lawrenceville, GA), 100 units/mL Penicillin, and 100 µg/mL Streptomycin (Cat. #: 15140122, Gibco). All cells were cultured in T75 flasks (Cat. #: 10062-860, Corning, Lowell, MA, USA), which were placed in a 5% carbon dioxide, 37 ºC, and 100% humidity incubator. Cells were passaged every 3-4 days and harvested for experiments when the cell culture reached 70-90% confluency. MDA-MB-231 cells with 20 or fewer passages were used.

### 2.2 Tumor spheroids

A specially designed array of microwells was used for making MDA-MB-231 spheroids [30]. Briefly, an array of 36 × 36 microwells was first patterned on a 1mm-thick agarose gel membrane using a soft lithography method. Each microwell is cylindrical in shape with a diameter of 200 µm and a depth of 250 µm (See Fig. S1). The agarose gel surface provides low adhesion surfaces to the cells, making it easier for the cells to cluster together and form spheroids. One microwell array was then placed in each well of a 12-well plate (Cat. #: 07- 200-82, Corning). Within each well, we placed 1.25 million MDA-MB-231 cells suspended in 2.5 mL of DMEM/F12 media (DMEM/F-12, Cat. #: 11320033, Gibco) supplemented with 5% horse serum (Cat. #: S12150, Gibco), 5% EGF (Cat. #: PHG0311, Gibco), 0.5 mg/mL hydrocortisone (Cat. #: H0888- 1G, Sigma-Aldrich), 100 ng/mL cholera toxin (Cat #: C-8052, Sigma), 10 µg/mL insulin (Cat. #: I1882, Sigma), 100 units/mL Penicillin, and 100 µg/mL Streptomycin (Cat. #: 15140122, Invitrogen). The 12-well plate was then kept in an incubator (Forma, Thermo Scientific, Asheville, NC, USA) at 37 ºC, 5% carbon dioxide, and 100% humidity for 5 days before harvesting. On day 3, the medium was changed to fresh medium. The spheroids were starved in DMEM supplemented with 1% FBS for 8 hours before harvesting to enhance their response to the EGF gradient. The average diameter of the spheroids was about 100 µm at the time of collection. For each experiment, the spheroids were collected from 6 microwell arrays, and a Falcon® Cell Strainer (Cat. #: 352350, Corning) with 70 µm pores was used to collect the spheroids. More details of the spheroid making process can be found in Ref. [31].

### 2.3 3D spheroid culture

To make 3D spheroid cultures, we suspended tumor spheroids in Type I collagen (rat tail tendon Cat. #: 354249, Corning). The collagen stock was first diluted to 5 mg/mL with 0.1% acetic acid and stored at 4 °C before the experiments. For each experiment, we typically prepared 200 μL tumor spheroid embedded collagen with a collagen concentration of 1.5 mg/mL, and a spheroid concentration of about 7,800 spheroids/mL. To do this, a 60 μL collagen stock (5 mg/mL) was first titrated with 1.32 μL 1 N NaOH and 20 μL 10X M199 (Cat. #: M0650-100ML, Sigma) to yield a final pH of ~7.4. Then, 118.7 μL of spheroids culture with DMEM supplemented with 1% FBS was added to reach a final volume of 200 μL. This stock was placed in ice until it was introduced into the microfluidic channel.

*2.4 Microfluidic device setup*

*Surface activation.* A standard 1"× 3" glass slide (Fisher Scientific, Pittsburgh, PA) was first treated with 1% PEI for 10 minutes. After rinsing it with sterile dH$_2$O, it was treated with glutaraldehyde for 30 minutes. The glass slides were left in a biohood overnight in sterile dH$_2$O. The glass slides were then washed with sterile dH$_2$O and dried before use. Surface activation is a crucial step because it prevents the collagen matrices from detaching from the glass slide due to the cell generated traction forces.

*Device assembly.* With a 1-mm thick polycarbonate spacer placed around the device pattern of the master silicon wafer, 2.5% boiled agarose gel was poured over the pattern. The agarose gel was pressed with a standard 1" × 3" glass slide to form a 1 mm-thick agarose gel. After allowing the gel to cool down to room temperature, the gel was then lifted off the pattern gently. The holes for the inlets and the outlets were made with a 2-mm biopsy punch (Cat. #: 21909-132, Miltex Inc. York, PA), and then the gel was submerged in L-15 medium (Cat. #: 11415064, Gibco) with 2.5% FBS, 100 units/mL Penicillin, and 100 μg/mL Streptomycin (Cat. #: 15140122, Gibco) for an hour. The agarose gel membrane with the spacer around was placed on a standard 1" × 3" glass slide, which was then sandwiched between a Plexiglas manifold and a stainless-steel frame (See Fig. 1). All parts except for the Plexiglas manifold and the polycarbonate spacer were autoclaved for sterility, and all channels were primed with L-15 media. In order to prevent evaporation, all the inlets and outlets were then plugged until media or spheroid-mixed collagen gel was introduced.

*3D spheroid seeding.* 15 μL of well mixed spheroid-embedded collagen was first pipetted slowly into the center channel of the device with an ice pack placed underneath. After plugging the device to prevent evaporation, the device was then incubated at 37 °C and 5% CO$_2$ for 45 minutes for collagen polymerization. To prevent the spheroids from settling down at the bottom during the collagen polymerization process, the device was first placed upside-down, where the glass slide was on top. Then, the device was flipped a total of three times at time points 11, 16, and 30 minutes. Using this protocol, most of the spheroids were located in the middle z plane of the center channel.

*Flow control and EGF gradient generation.* The flows into the two side channels were provided by two 10 mL syringes (Cat. #: 303134, Becton Dickinson, Franklin Lakes, NJ, USA) pumped with a syringe pump (KDS230, KD Scientific, Holliston, MA) through a medical grade tubing (Cat. #: AY202431-CP, Cole-Parmer, Vernon Hills, IL). All fluids from the syringes passed through a 0.2 μm filter (Cat. #: MS-3301, Pall Corporation, Port Washington, NY) before going into the inlets of the device to prevent air bubbles from entering the microfluidic channels. L-15 medium supplemented with 2.5% FBS was pumped through the sink channels, and the same medium supplemented with 8.33 nM EGF (Cat #: 354052, Corning, Lowell, MA, USA) was pumped through the source channels at a flow rate of 1 μL/min to generate the EGF gradient. For control, both the sink channel and the source channel were pumped with L-15 medium supplemented with 2.5% FBS.

*2.5 Imaging and data analysis*

All images were taken with a 10X magnification objective lens (NA = 0.25, Olympus America, Center Valley, PA, USA) on an inverted epi-fluorescent microscope (IX 81, 40 Olympus America, Center Valley, PA, USA) and a CCD camera (ORCA-R2, Hamamatsu Photonics, Bridgewater, NJ, USA). The light source for fluorescence imaging was provided by the X-Cite series 120PC unit (Excelitas Technologies, Waltham, MA, USA). The scope was surrounded by a stage incubator (Precision Plastics Inc., Beltsville, MD, USA) that maintained a temperature of 37 °C, humidity of ~70%, and atmospheric CO$_2$ level since L-15 medium was used. The device was placed on an automated x-y microscope stage (MS-2000, Applied Scientific Instrumentation, Eugene, OR), and images were taken every 10 minutes for 48 hours using CellSens software (Olympus America, Center Valley, PA, USA). In each experiment, bright field and fluorescence images were taken at 12 selected positions at each time point. The time point 0 is defined as the time when the imaging started, which is about 1 hour after the introduction of the spheroid embedded collagen to the center channel.

The spheroid invasion was characterized by analyzing the area and the center of mass of the spheroid core over time. Here, we define spheroid core as the center part of the spheroid where the cells are still interconnected (See the yellow outlines Fig. S2). The spheroid core was tracked using the fluorescence images of the MDA-MB-231 spheroid (Fig. S2) with a particle detection module in ImageJ (National Institute of Health, Bethesda, MD). For each time-series image, all pixels with intensity greater than 400 (the mean intensity of all the images is 333) were selected, of which particles greater than 7000 μm² were segmented. The area and the center of mass of the segment were then calculated using ImageJ. The area of the spheroid expansion was normalized by the initial spheroid area (Control: 8513 ± 1139 μm², EGF: 10984 ± 896 μm²). Note that the spheroid area is measured in the x-y plane. The directionality of the spheroid movement was characterized by displacement of the spheroid center of mass along the gradient direction with respect to the initial position at t=0. Here, the center of mass is the brightness-weighted average of the x and y coordinates of the spheroid core.

Cell trajectories were obtained using the manual tracker in ImageJ (National Institutes of Health) using the time-lapse images. Single-cell migration parameters, speed, velocity, persistence length, and mean squared displacement (MSD) were calculated using these trajectories [23]. The cell speed was defined as the total length of each track divided by the time duration. The cell velocity was defined as the distance between the cell starting and ending locations divided by the time duration. The cell persistence length was defined as the distance between the cell starting and ending locations divided by the length of the cell trajectory. MSD was defined as the average of the distance traveled between neighboring time points squared. To minimize experiment-to-experiment variability, data acquired were normalized for each experiment. The cell speed in the EGF condition is normalized by the average cell speed of the control condition from each experiment (0.182, 0.067 μm/min), and the cell velocity along the direction of gradient (x-velocity) is normalized by first subtracting the x-velocity from the average cell velocity along the x-direction of the control group (0.011, -0.003 μm/min) then dividing by the average speed of the cells in the control condition (0.182, 0.067 μm/min). The persistence length of the cells in the EGF condition is normalized by the average persistence of the cells in the control condition (0.46, 0.51), and the persistence length of the cells in the EGF condition along the gradient direction (x-plength) is normalized by subtracting the average persistence length along the gradient direction of control (0.03, -0.05). For calculation of aspect ratio, the ellipse function from ImageJ was used to fit a single cell. Then, the ratio of major over minor axis was used to obtain the aspect ratio values. Cells with aspect ratio less than 2 were considered amoeboid and greater than 2 mesenchymal [32, 33]. Non-parametric t-test (Mann-Whitney test) was carried out using Prism (GraphPad Software, Inc., La Jolla, CA).

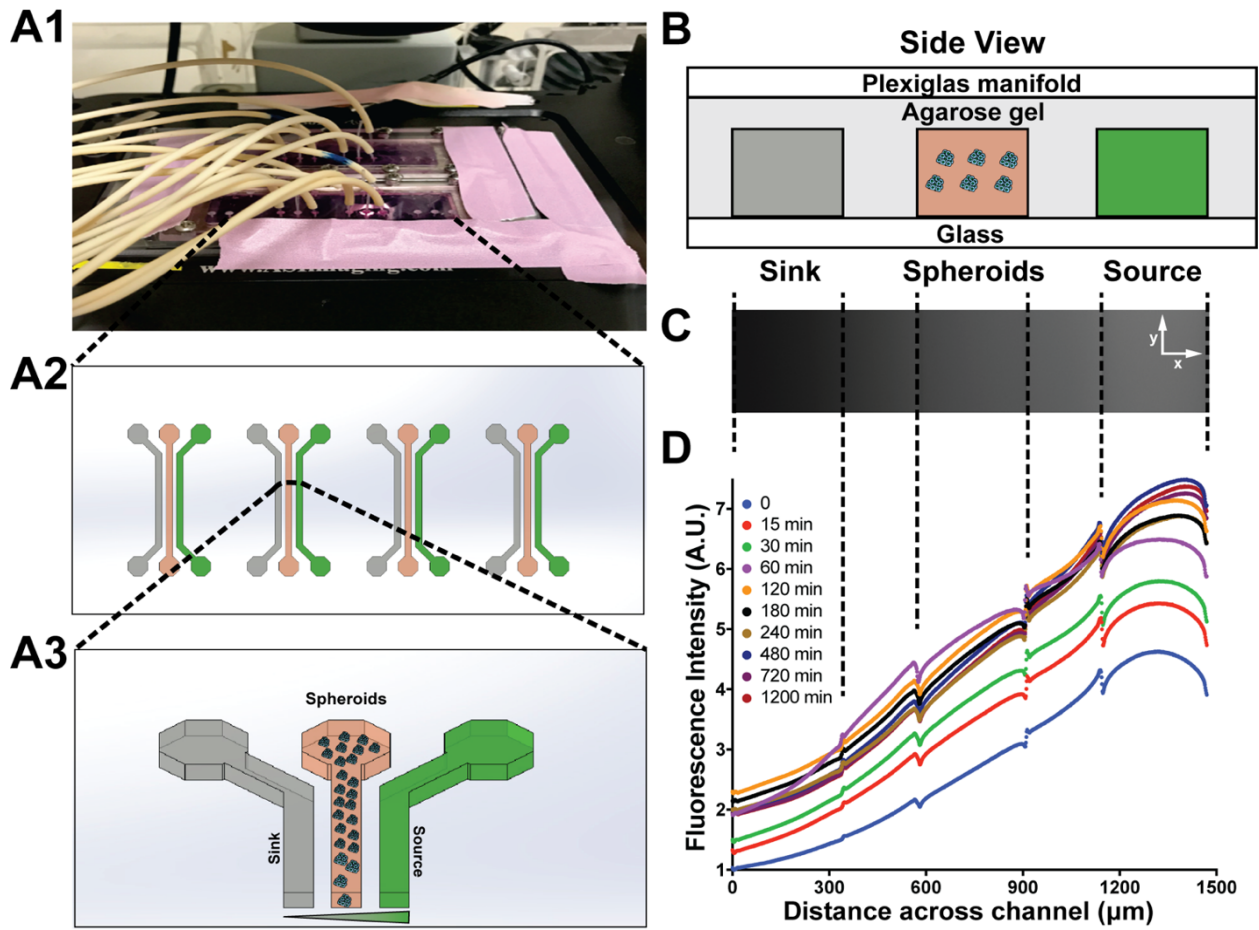

**Figure 1: Microfluidic setup for tumor chemotaxis experiments and device calibration.** (A1) An image of a microfluidic device for 3D tumor spheroid chemotaxis studies. Two functional chips are placed on a microscope stage for multiple position imaging. (A2-3) Designing principles of a microfluidic gradient generator. Four three-channel devices were patterned on a 1mm thick agarose gel membrane placed on top of a 1" x 3" microscope slide (A2). Each device consists of three parallel channels. Spheroid-embedded collagen was introduced in the middle channel. EGF and media were introduced into source and sink channels, respectively (A3). All channels have a width of 330 μm and a depth of 250 μm, and the distances between adjacent channels are 240 μm each. (B) A cross section view of the three-channel device. Drawing not to scale. (C) A fluorescent image of all three channels at a steady state, t = 60 min. (D) Gradient generation characterization. Time evolution of FITC dextran concentration profiles revealed by the fluorescence images across three channels. It takes about 1 hours to reach a steady state. The gradient profile can be kept as long as the side channels are pumped continuously.

## 3. Results and Discussion

### 3.1 Microfluidic experimental setup, EGF gradient generation and calibration.

A hydrogel based microfluidic chemical gradient generator previously developed in our lab was adapted here for tumor spheroid chemotaxis experiments (See Fig. 1) [19, 34]. Briefly, a pattern of four three-channel devices was imprinted in a 1-mm thick agarose gel and sandwiched between a Plexiglas manifold and a stainless-steel frame (See Fig. 1A, B). MDA-MB-231 tumor spheroids embedded in type I collagen at a final concentration of 1.5 mg/mL were introduced into the middle channel. After collagen polymerization, L-15 medium with 2.5% FBS with or without 8.33 nM EGF was introduced to the source and sink channels, respectively. This is comparable to the reported apparent $K_d$ (2-15 nM) of EGFR towards its ligands [35, 36]. As a result, a linear EGF gradient of 5.14 nM/mm was established in the middle channel via molecular diffusion through the agarose gel, resulting in about 0.514 nM EGF concentration difference from the front and the back end of the spheroid in the EGF gradient.

To verify that the microfluidic device can establish and maintain a steady EGF (MW: 6400 Da) gradient across the center channel for a prolonged period of time (~ 24 hours), we characterized the gradient generation using FITC dextran (MW: 4000 Da). Here, a 0.1 mM FITC dextran solution was pumped through the source channel, and blank PBS was pumped through the sink channel at a flow rate of 1 μL/min. Time-lapse fluorescence images of all three channels were taken every minute for over 24 hours. We define t = 0 to be the time when the syringe pump was started right after an initial flush of the fluids in their respective channels to remove all air bubbles within the channels. The fluorescence concentration profiles in Fig. 1D show that it takes about 1 hour to reach a steady state. This is consistent with the theoretical calculation using a first order approximation for the establishment time, ~ $L^2/2D$, where L is the distance between the two side channels (0.81 mm), and D is the diffusion coefficient of FITC dextran (94.77 μm$^2$/s). The gradient profile was kept steady for more than 24 hours in our experiments as long as the flows in the side channels were maintained. The gradient of 5.14 nM/mm was calculated based on the fluorescence signal across the channel from the calibration. This is lower than the expected gradient of 10.28 nM/mm. We conjecture that this difference arises due to the diffusion into the surrounding agarose gel.

### 3.2 An EGF gradient promoted spheroid core spreading and mild chemotaxis into the 3D ECM

Tumor spheroids were shown to spread out significantly more in the presence of EGF gradients than in the absence of EGF within the 48 hours observation time (See Fig. 2A and Fig. S2, and Movie S1,2). When examining the movies of tumor spheroid invasion carefully, tumor spheroids were seen to initially spread out, and then individual cells started to detach from the spheroid core and invaded into the ECM. We define the spheroid core as the main tumor body where all the tumor cells are connected with each other. In the following, we quantify the spheroid invasion processes in two steps: (i) the spheroid core dynamics; (ii) motility of the cells that invaded out of the spheroid core.

For spheroid core dynamics, we examined the area and the center of mass of the spheroid core as a function of time. Fig. 2B shows that the area of the spheroid core spread out much more in presence of EGF than those in absence of EGF. Interestingly, spheroid spreading leveled off at around t = 30 hours in presence of the EGF gradient, whereas in control, the spheroids were spreading throughout the entire 48 hours. This is likely caused due to the spheroid core making contact with the microfluidic channel boundary starting at around t = 30 hours in presence of the EGF gradient.

When examining the displacement of the center of mass of the spheroid core as a function of time, we found a mild chemotaxis along the EGF gradient. In Fig. 2C, a significant deviation in x-displacement of the center of mass of the spheroid core from its initial position appeared after 16 hours, and then leveled off when t ~ 30 hours. We conjecture that it takes time for the spheroid to respond to EGF, and this chemotactic response became less sensitive when the spheroid core started to make contact with the microfluidic channel boundary at around t = 30hrs. In the control, on the other hand, the center of mass of the spheroid core did not move significantly, indicating that there was no significant chemotactic response. The width of the channel can be made wider in future experiment to decrease the boundary influence of this experiment.

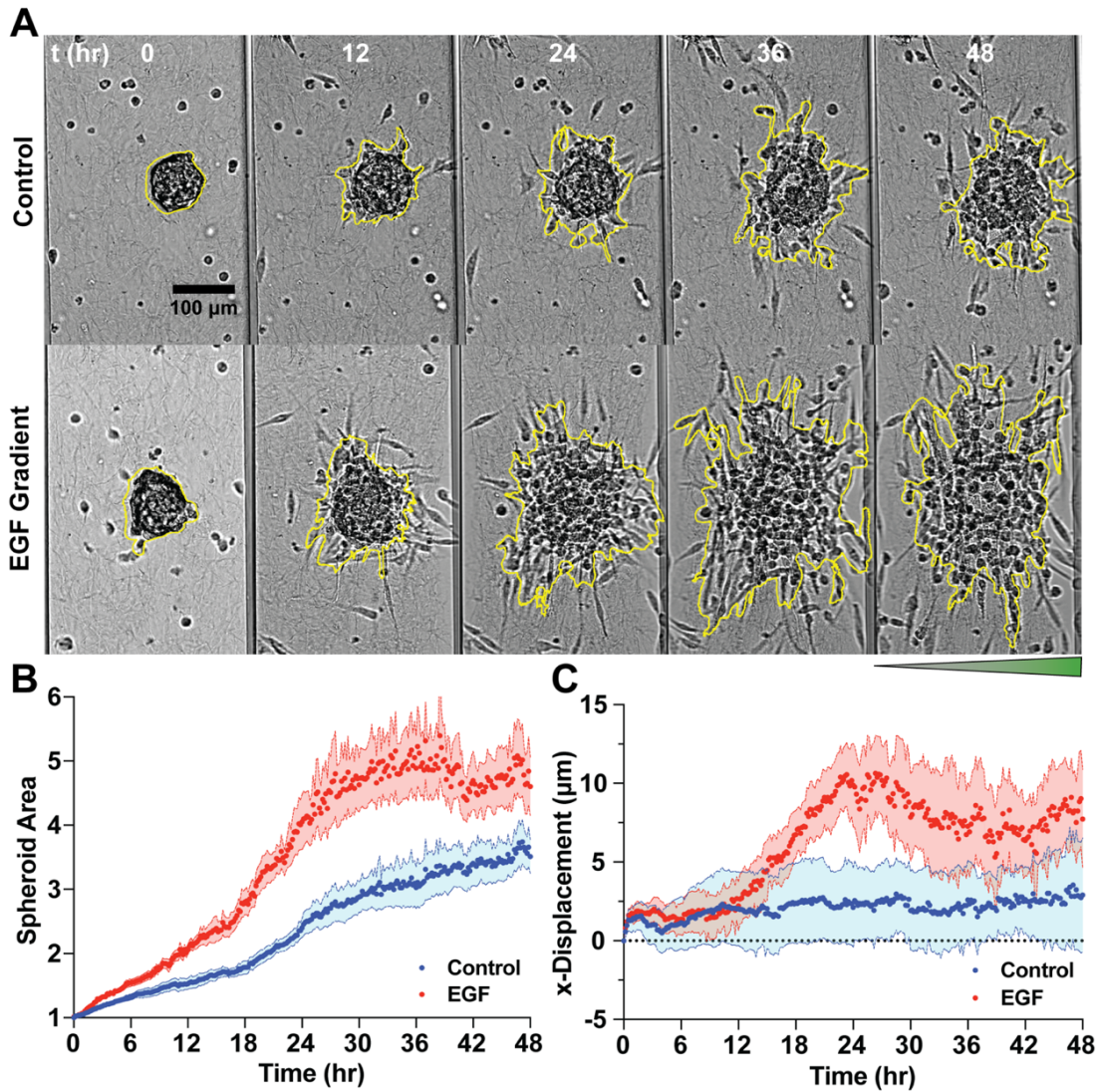

**Figure 2: EGF gradient promoted tumor spheroid invasion within a collagen matrix.** (A) Bright-field time lapse images of MDA-MB-231 spheroids embedded in collagen in the presence or absence of an EGF gradient. Using the particle detection module in ImageJ, we segmented the tumor cells that are still connected with each other (marked by the yellow outline) and named it the spheroid core. The segmentation is carried out using the fluorescence images, and the outlines are superimposed onto the bright field images as shown here. The EGF gradient (5.14 nM/mm) was imposed along the x-direction. The tumor spheroids were embedded within a 1.5 mg/mL collagen matrix. (B) Normalized area of the spheroid core in the presence and absence of EGF gradient. (C) The displacement of the center of mass of the tumor spheroid core with respect to its initial position along the gradient direction. In B and C, the time between two consecutive data points was 10 minutes. The blue dots are experimental results from the control case, and the red dots are those from the EGF gradient case. The lighter shades represent the SEM for each case. A total of 4 spheroids and 5 spheroids were analyzed for control and EGF gradient case, respectively.

One interesting phenomenon that we observed during the experiment was that often tumor spheroids generated enough cell traction force to detach collagen from its boundaries. Despite the application of the surface treatment process on the glass slide to prevent collagen gel detachment, the force that the spheroids exerted on the collagen gel was sufficient to detach the collagen from the agarose gel wall for most spheroids (Fig. S3). Also seen in Fig. S3 is the fact that the

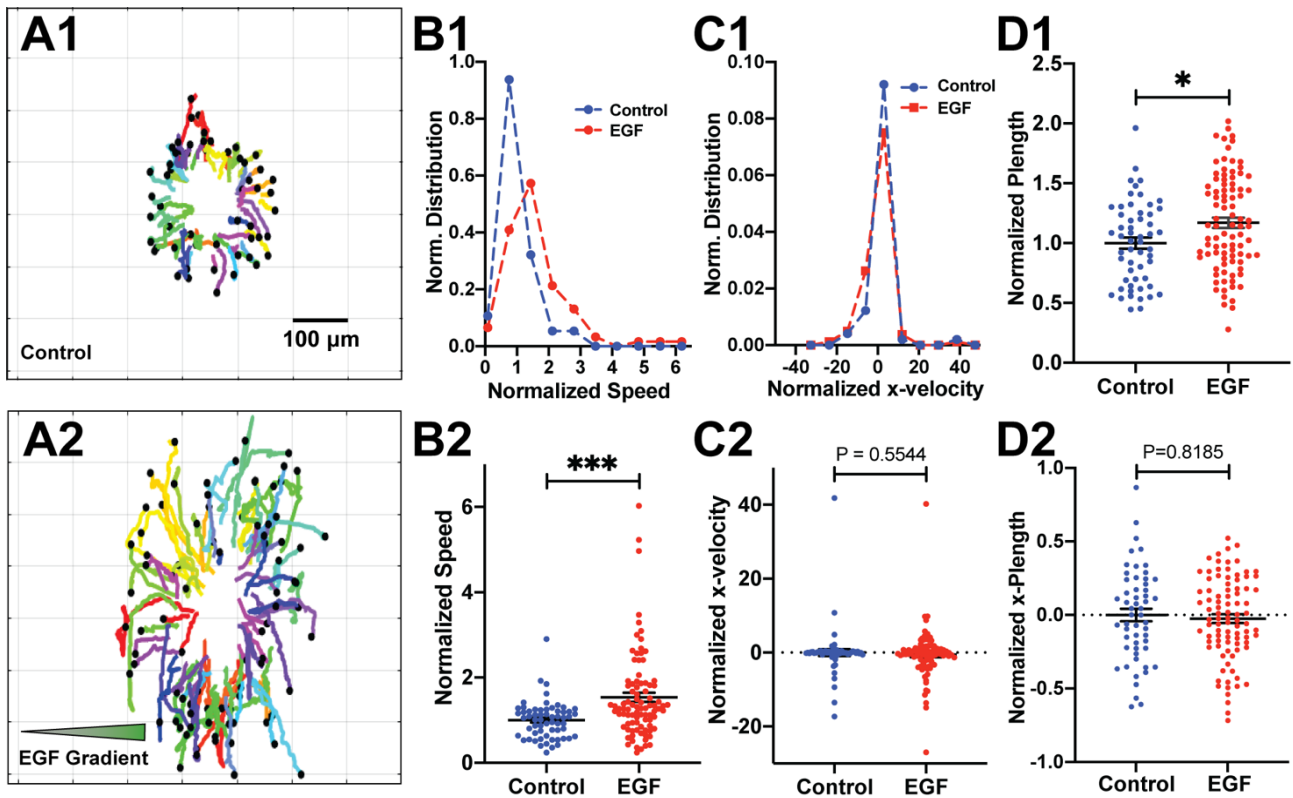

**Figure 3: Single tumor cells displays distinct chemokinesis but no obvious chemotactic behavior in EGF gradient**
(A) Trajectories of cells in control (A1) and in EGF gradient (A2) condition. Each colored line is a cell trajectory, and cells were tracked after they detached from the spheroid. 55 and 90 cells were tracked in (A1) and (A2), respectively, and the track duration was between 3 - 45 hours with an average of 20.5 ± 10.0 hours for the EGF gradient case, and 2 - 45 hours with an average of 21.8 ± 11.1 hours for control. The EGF gradient was generated by flowing 8.33 nM EGF and medium in source and sink channel, respectively. (B1) Distribution of normalized cell migration speed in control versus in EGF gradient. (B2) Normalized cell migration speed. ****: p = 0.0002. (C1) Distribution of normalized velocity Vx along the EGF gradient direction. (C2) Normalized cell velocity along the direction of gradient. (D1) Normalized cell migration persistence. *: p = 0.0104. (D2) Normalized cell migration persistence along the direction of gradient. Results from (B) - (D) were computed from trajectories shown in (A1) and (A2). The stars were obtained using a nonparametric t-test compared to the control group (Mann-Whitney test with ****: P < 0.0001, ***: P < 0.001, **: P detachment area is larger in the EGF case than its counterpart without EGF, indicating that tumor spheroids generate more traction force in the presence of EGF. In our experiments, only the spheroids that did not detach the collagen gel were included in our data analysis because the center of mass of the spheroids were seen to have a corresponding shift with the gel peeling. This yielded a total of 4 spheroids for the control condition and a total of 5 spheroids for the EGF gradient condition for the data analysis shown in Fig. 2 (A-C). These data were combined from two independent experiments.

### 3.3 The EGF gradient promoted motility of the cells that invaded away from the spheroid core

The motility of the cells that invaded away from the spheroid core was examined using the bright field time-lapse image series as those shown in Fig. 2A. Cell trajectories of each migrating cell in the presence/absence of the EGF gradients were obtained using manual tracker in ImageJ (See Fig. 3A1, A2). Note that the starting point of each cell track was when the cell first detached from the spheroid core. Clearly, cell motility was enhanced in presence of the EGF gradient when compared to the control condition (Fig. 3 A1-A2). It is interesting to note that the cell trajectories in Fig. 3A displayed an alignment along the y or channel direction. This was more evident in presence of the EGF gradient than without. This phenomenon was caused by the collagen alignment along the channel due to flow shear stress when unpolymerized collagen was introduced into the narrow channel as reported in early work [37].

We computed cell speed, velocity and persistence using the cell trajectories shown in Fig. 3A. Distributions of cell speed as well as their average speed (Fig. 3B1, B2) show that cells are much more motile in presence of the EGF gradient than in absence of the EGF gradient. The average normalized speed of the cells is 1.54 in the presence of EGF, in contrast to 1.00 in the absence of EGF, yielding a 54% increase. To evaluate whether the cells that invaded away from the spheroid core are chemotactic to the EGF gradient, the distributions of cell x-velocity (Fig. 3C1) and average x-velocity with and without EGF gradient were plotted (Fig. 3C2). Both plots (Fig. 3C1 and C2) showed that the detached cells did not show a distinguishable chemotactic response. The normalized persistence length is 1.17 in presence of the EGF gradient, in contrast to 1.00 in absence of EGF, which is an increase of 17% (Fig. 3D1). Again, there was no significant difference in the persistence of the detached cells along the direction of the EGF gradient in presence vs. absence of the EGF gradient (Fig. 3D2).

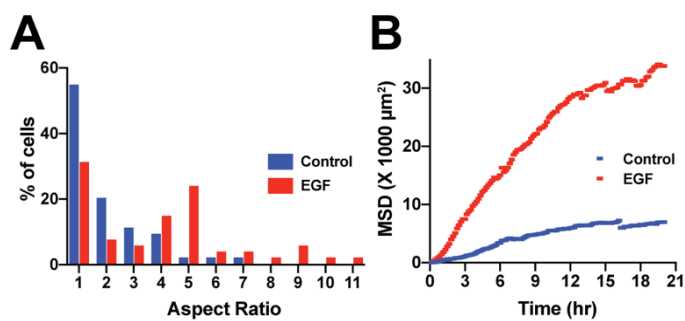

**Figure 4: EGF promoted mesenchymal cell morphology and cell invasion.** (A) Distribution of the aspect ratio of the tumor cells with and without EGF gradients. A total of 55 cells each were randomly selected from EGF and control case at t=48 hour. (B) Tumor cell mean square displacements (MSDs) for MDA-MB-231 cells with and without EGF. This result was computed using the tracks shown in Fig. 3 (A1)- (A2).

Previously, we reported a study on the role of EGF gradients in single cell migration within a 3D collagen gel at a concentration of 1.5 mg/mL [23]. The results presented in this study were consistent with our previous report in that chemokinesis was observed in the presence of the EGF gradient, but no chemotaxis for single cells invading in a collagen matrix. Interestingly, we did find that tumor cells were more sensitive to EGF in the spheroid assay in contrast to the single cell assay. We note that an overall 54% increase in speed was observed in the spheroid assay from this study, whereas only ~12% increase in speed was observed in the 3D single cell assay. It is also interesting to note that an overall ~17% increase in persistence length was observed in the spheroid assay, whereas only ~10% increase in persistence length was observed in the single-cell assay. We conjecture that the stronger traction force from the spheroids may significantly remodel the architecture of the ECM, which led to the differential sensitivity of single tumor cells to EGF gradients and a radial invasion pattern from the spheroid core rather than a random walk. This conjecture needs to be verified in future experiments where ECM architecture will be visualized in both cases.

### 3.4 EGF promoted a mesenchymal over amoeboid cell morphology and single cell spreading in space

We found that cells were more elongated and exhibited more mesenchymal over amoeboid cell motility in the presence of the EGF gradient. This is shown in the distribution of cell aspect ratios in presence/absence of the EGF gradient (Fig. 4A). In this analysis, we labeled a cell as amoeboid if the aspect ratio was less than 2 to be consistent with previous publications [32, 33].

To analyze the individual cells spreading in space, MSDs of the MDA-MB-231 tumor cells were computed using the cell trajectories (Fig. 4B). Though the experiments were done in 3D, the cells were tracked in 2D. Thus, a first-order approximate mean squared displacement equation, $MSD = 4Dt$, was used, where MSD is the average of the displacement squared of the cells at various time points, D is the diffusion coefficient of the cells, and t is time in minutes. The diffusion coefficients for the cells in presence and absence of the EGF gradient were 7.04 $\mu m^2/min$ and 1.45 $\mu m^2/min$, respectively. The drastic difference in the diffusion coefficients showed that EGF promoted cell spreading in space significantly.

### 4. Conclusion and Future Perspective

Using a 3D spheroid invasion assay, we observed a strong chemokinesis and a mild chemotactic response of tumor spheroids to EGF gradients. In contrast to previous chemotaxis studies where single cells were embedded within an ECM, tumor cells associated with tumor spheroids exhibited an enhanced sensitivity to the EGF gradients compared to single cells embedded within an ECM. This result highlights the importance of a tumor spheroid assay in performing chemotaxis studies, or for future drug testing.

Looking ahead, a number of important questions need to be addressed. First is the cell-ECM interaction. It will be interesting to follow ECM remodeling by the tumor spheroid traction force and study how the fiber alignment affects chemotaxis. Second is the roles of cell-ECM, cell-cell adhesion molecules in the chemotactic response. For cell-ECM interactions, we can monitor integrin expression as a function of EGF concentration. For cell-cell adhesion, we can use spheroids made of cell lines that have higher cell-cell adhesion molecules (E-cadherin) such as MCF-7. It is likely that these cell lines will be able to better mimic the behaviors presented in the theoretical studies on cell cluster migration [24-26]. Third is the inclusion of a natural control where the spheroids are exposed to a uniform EGF concentration, which could help isolating the effect of gradient sensing from the chemokinetic effect of EGF.

**References**

1. Roussos ET, Condeelis JS, Patsialou A. Chemotaxis in cancer. Nat Rev Cancer. 2011;11(8):573-87.
2. Wu M, Swartz MA. Modeling Tumor Microenvironments In Vitro. Journal of Biomechanical Engineering. 2014;136(2):7.
3. Chambers AF, Groom AC, MacDonald IC. Dissemination and growth of cancer cells in metastatic sites. Nat Rev Cancer. 2002;2(8):563-72.
4. Zhou ZN, Sharma VP, Beaty BT, Roh-Johnson M, Peterson EA, Van Rooijen N, et al. Autocrine HBEGF expression promotes breast cancer intravasation, metastasis and macrophage-independent invasion in vivo. Oncogene. 2014;33(29):3784-93.
5. Goswami S, Sahai E, Wyckoff JB, Cammer M, Cox D, Pixley FJ, et al. Macrophages promote the invasion of breast carcinoma cells via a colony-stimulating factor-1/epidermal growth factor paracrine loop. Cancer research. 2005;65(12):5278-83.
6. Berger MS, Greenfield C, Gullick WJ, Haley J, Downward J, Neal DE, et al. Evaluation of epidermal growth factor receptors in bladder tumours. Br J Cancer. 1987;56(5):533-7.
7. Gullick WJ. Prevalence of aberrant expression of the epidermal growth factor receptor in human cancers. Br Med Bull. 1991;47(1):87-98.
8. Lemoine NR, Hughes CM, Gullick WJ, Brown CL, Wynford-Thomas D. Abnormalities of the EGF receptor system in human thyroid neoplasia. Int J Cancer. 1991;49(4):558-61.
9. Libermann TA, Nusbaum HR, Razon N, Kris R, Lax I, Soreq H, et al. Amplification, enhanced expression and possible rearrangement of EGF receptor gene in primary human brain tumours of glial origin. Nature. 1985;313(5998):144-7.
10. Salomon DS, Brandt R, Ciardiello F, Normanno N. Epidermal growth factor-related peptides and their receptors in human malignancies. Crit Rev Oncol Hematol. 1995;19(3):183-232.
11. Tillotson JK, Rose DP. Endogenous secretion of epidermal growth factor peptides stimulates growth of DU145 prostate cancer cells. Cancer Lett. 1991;60(2):109-12.
12. Rauch J, Volinsky N, Romano D, Kolch W. The secret life of kinases: functions beyond catalysis. Cell Communication and Signaling. 2011;9(1):23.
13. Wang Y, Deng W, Zhang Y, Sun S, Zhao S, Chen Y, et al. MICAL2 promotes breast cancer cell migration by maintaining epidermal growth factor receptor (EGFR) stability and EGFR/P38 signalling activation. Acta Physiol (Oxf). 2018;222(2).
14. Seshacharyulu P, Ponnusamy MP, Haridas D, Jain M, Ganti AK, Batra SK. Targeting the EGFR signaling pathway in cancer therapy. Expert Opin Ther Targets. 2012;16(1):15-31.
15. Zigmond SH. Ability of polymorphonuclear leukocytes to orient in gradients of chemotactic factors. The Journal of Cell Biology. 1977;75(2):606.
16. Boyden S. The chemotactic effect of mixtures of antibody and antigen on polymorphonuclear leucocytes. J Exp Med. 1962;115(3):453-66.
17. Jeon NL, Dertinger SKW, Chiu DT, Choi IS, Stroock AD, Whitesides GM. Generation of Solution and Surface Gradients Using Microfluidic Systems. Langmuir. 2000;16(22):8311-6.
18. Kim BJ, Wu M. Microfluidics for Mammalian Cell Chemotaxis Annals of Biomedical Engineering. 2012;40(6):1316-27.
19. Haessler U, Kalinin Y, Swartz MA, Wu M. An agarose-based microfluidic platform with a gradient buffer for 3D chemotaxis studies. Biomedical microdevices. 2009;11(4):827-35.
20. Haessler U, Pisano M, Wu M, Swartz MA. Dendritic cell chemotaxis in 3D under defined chemokine gradients reveals differential response to ligands CCL21 and CCL19. Proc Natl Acad Sci U S A. 2011;108(14):5614-9.
21. Abhyankar VV, Toepke MW, Cortesio CL, Lokuta MA, Huttenlocher A, Beebe DJ. A platform for assessing chemotactic migration within a spatiotemporally defined 3D microenvironment. Lab Chip. 2008;8(9):1507-15.
22. Wang S-J, Saadi W, Lin F, Nguyen CM-C, Jeon NL. Differential effects of EGF gradient profiles on MDA-MB-231 breast cancer cell chemotaxis. Experimental cell research. 2004;300(1):180-9.
23. Kim BJ, Hannanta-anan P, Chau M, Kim YS, Swartz MA, Wu M. Cooperative roles of SDF-1alpha and EGF gradients on tumor cell migration revealed by a robust 3D microfluidic model. PLoS One. 2013;8(7):e68422.
24. Camley BA. Collective gradient sensing and chemotaxis: modeling and recent developments. J Phys Condens Matter. 2018;30(22):223001.
25. Varennes J, Han B, Mugler A. Collective Chemotaxis through Noisy Multicellular Gradient Sensing. Biophysical Journal. 2016;111(3):640-9.
26. Gopinathan A, Gov NS. Cell cluster migration: Connecting experiments with physical models. Seminars in Cell & Developmental Biology. 2019;93:77-86.
27. Debets VE, Janssen LMC, Storm C. Enhanced persistence and collective migration in cooperatively aligning cell clusters. Biophysical Journal. 2021;120(8):1483-97.
28. Ellison D, Mugler A, Brennan MD, Lee SH, Huebner RJ, Shamir ER, et al. Cell–cell communication enhances the capacity of cell ensembles to sense shallow gradients during morphogenesis. Proceedings of the National Academy of Sciences. 2016;113(6):E679-E88.
29. Hwang PY, Brenot A, King AC, Longmore GD, George SC. Randomly Distributed K14+ Breast Tumor Cells Polarize to the Leading Edge and Guide Collective Migration in Response to Chemical and Mechanical Environmental Cues. Cancer Research. 2019;79(8):1899-912.

30. Song W, Tung CK, Lu YC, Pardo Y, Wu M, Das M, et al. Dynamic self-organization of microwell-aggregated cellular mixtures. Soft Matter. 2016;12(26):5739-46.
31. Lu Y-C, Song W, An D, Kim BJ, Schwartz R, Wu M, et al. Designing compartmentalized hydrogel microparticles for cell encapsulation and scalable 3D cell culture. Journal of Materials Chemistry B. 2015;3(3):353-60.
32. Huang YL, Tung C-k, Zheng A, Kim BJ, Wu M. Interstitial flows promote amoeboid over mesenchymal motility of breast cancer cells revealed by a three dimensional microfluidic model. Integrative Biology. 2015;7(11):1402-11.
33. Petrie RJ, Gavara N, Chadwick RS, Yamada KM. Nonpolarized signaling reveals two distinct modes of 3D cell migration. Journal of Cell Biology. 2012;197(3):439-55.
34. Cheng S-Y, Heilman S, Wasserman M, Archer S, Shuler ML, Wu M. A hydrogel-based microfluidic device for the studies of directed cell migration. Lab on a Chip. 2007;7(6):763-9.
35. Macdonald-Obermann JL, Pike LJ. Allosteric regulation of epidermal growth factor (EGF) receptor ligand binding by tyrosine kinase inhibitors. The Journal of biological chemistry. 2018;293(35):13401-14.
36. Schlessinger J. The epidermal growth factor receptor as a multifunctional allosteric protein. Biochemistry. 1988;27(9):3119-23.
37. Lee P, Lin R, Moon J, Lee LP. Microfluidic alignment of collagen fibers for in vitro cell culture. Biomed Microdevices. 2006;8(1):35-41.

# Supporting Materials

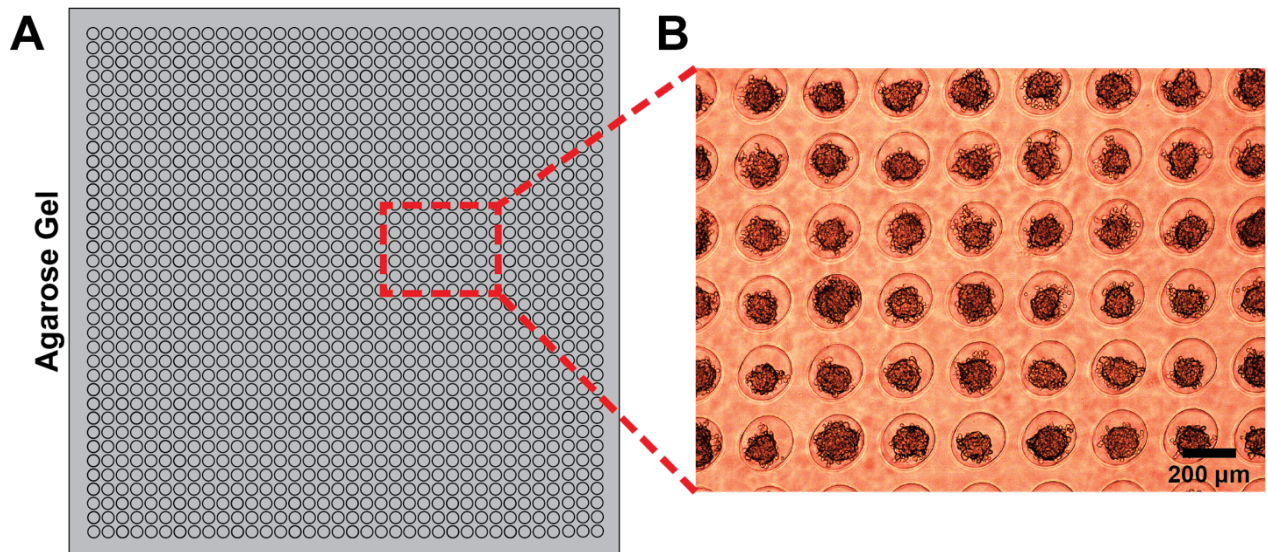

**Figure S1. Schematic of a microwell array platform for MDA-MB-231 spheroid formation.** (A) A 36 x 36 microwell array on an agarose gel was imprinted using soft lithography. Each microwell is cylindrical in shape with a diameter of 200 μm and a depth of 250 μm. (B) A 4X brightfield image of agarose microwell array with MDA-MB-231 spheroids taken at day 5.

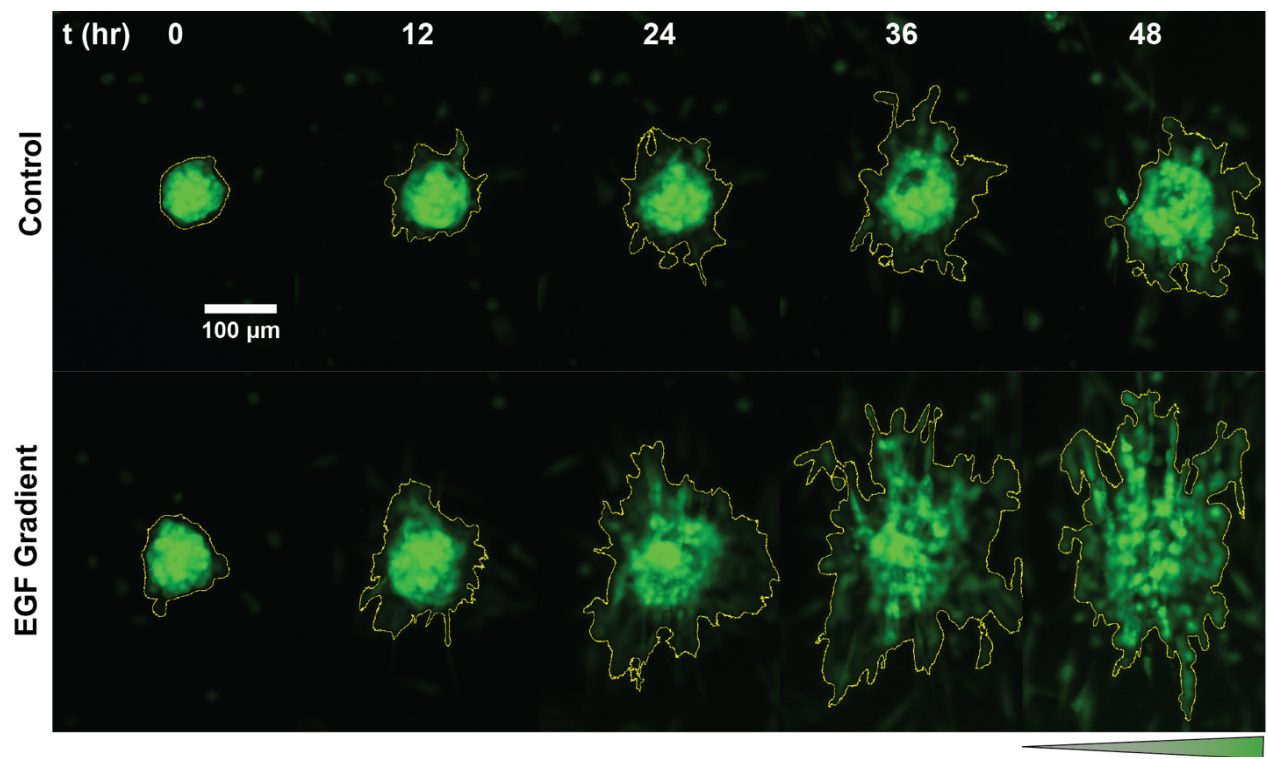

**Figure S2: Fluorescence time lapse images of MDA-MB-231 spheroids.** Fluorescence time lapse images of MDA-MB-231 spheroids embedded in collagen in the presence or absence of EGF gradients. The spheroids in these fluorescence images are the same spheroids as the ones shown in Fig.2A. Using the particle detection module in ImageJ, we segmented the tumor cells that are still connected with each other (marked by the yellow outline) and named it the spheroid core. The EGF gradient (5.14 nM/mm) was imposed along the x-direction. The tumor spheroids were embedded within a 1.5 mg/mL collagen matrix.

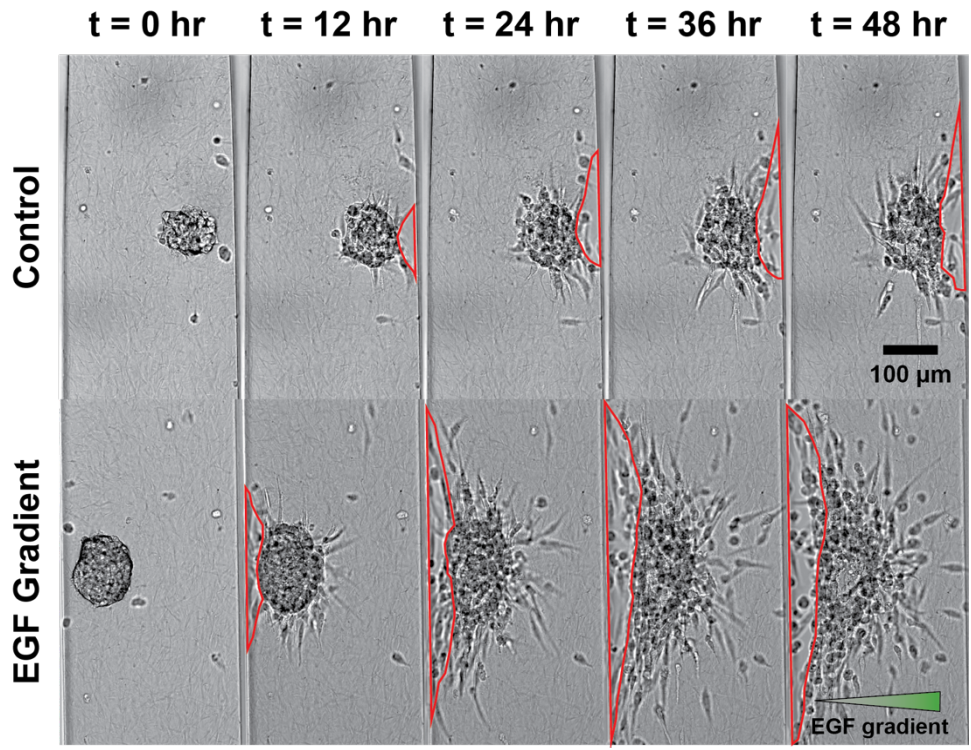

**Figure S3: Tumor spheroids exhibit large enough forces to peel off collagen.** Time-lapse images of MDA-MB-231 spheroid invasion are shown for 48 hours with and without an EGF gradient. The EGF gradient, 5.14 nM/mm, was imposed along the x-direction. The tumor spheroids were embedded within a 1.5 mg/mL collagen matrix. The scale bar represents 100 μm. MDA-MB-231 spheroid pulled the collagen and peeled it off from the glass slide and the agarose gel. The peeling region is marked by the red outline.

**Movie S1:** Fluorescence movie of MDA-MB-231 spheroid invasion is shown for 48 hours in EGF gradient. The EGF gradient, 5.14 nM/mm, was imposed along the x-direction. The tumor spheroids were embedded within a 1.5 mg/mL collagen matrix. The movie is 334.11 μm by 660.48 μm.

**Movie S2:** Fluorescence movie of MDA-MB-231 spheroid invasion is shown for 48 hours in the absence of EGF. The tumor spheroids were embedded within a 1.5 mg/mL collagen matrix. The scale bar represents 100 μm. The movie is 334.11 μm by 660.48 μm.